\documentclass[submission,copyright,creativecommons,noderivs,noncommercial]{eptcs}
\providecommand{\event}{International Workshop on Interactions, \\Games and Protocols (iWIGP 2011)} 

\usepackage[latin1]{inputenc}
\usepackage{amsmath}
\usepackage{breakurl}
\usepackage{underscore}
\newcommand{\newterm}{\emph}
\newcommand{\Unbeast}{\textsc{Unbeast}}
\newcommand{\Lily}{\textsc{Lily}}
\newcommand{\Acacia}{\textsc{Acacia}}
\newcommand{\Anzu}{\textsc{Anzu}}

\newcommand{\WringCite}{\textsc{Wring} \cite{DBLP:conf/cav/SomenziB00}}
\newcommand{\UnbeastCite}{\textsc{Unbeast} \cite{UnbeastTool,DBLP:conf/cav/Ehlers10}}
\newcommand{\LilyCite}{\textsc{Lily} \cite{LilyTool,DBLP:conf/fmcad/JobstmannB06}}
\newcommand{\AcaciaCite}{\textsc{Acacia} \cite{AcaciaTool,DBLP:conf/cav/FiliotJR09,DBLP:conf/atva/FiliotJR10}}
\newcommand{\AnzuCite}{\textsc{Anzu} \cite{DBLP:conf/cav/JobstmannGWB07}}
\newcommand{\AnzuExtCite}{\textsc{Anzu} \cite{AnzuTool,DBLP:conf/cav/JobstmannGWB07,DBLP:conf/date/BloemGJPPW07,DBLP:journals/entcs/BloemGJPPW07}}
\newcommand{\RatsyCite}{\textsc{Ratsy} \cite{DBLP:conf/cav/BloemCGHKRSS10}}

\newcommand{\CUDD}{\textsc{CuDD}} 
\newcommand{\LTLtoBACite}{\textsc{ltl2ba} \cite{DBLP:conf/cav/GastinO01}}
\newcommand{\SpotCite}{\textsc{spot}'s \textsc{ltl2tgba} \cite{DBLP:conf/mascots/Duret-LutzP04}}
\newcommand{\AP}{\mathsf{AP}}

\title{Experimental Aspects of Synthesis}
\def\titlerunning{Experimental aspects of synthesis}
\author{Rüdiger Ehlers
\institute{Reactive systems group, Saarland University}
}
\def\authorrunning{R. Ehlers}
\begin{document}
\maketitle

\begin{abstract}
We discuss the problem of experimentally evaluating linear-time temporal logic (LTL) synthesis tools for reactive systems.
We first survey previous such work for the currently publicly available synthesis tools, and then draw conclusions by deriving useful schemes for future such evaluations. 

In particular, we explain why previous tools have incompatible scopes and semantics and provide a framework that reduces the impact of this problem for future experimental comparisons of such tools. Furthermore, we discuss which difficulties the complex workflows that begin to appear in modern synthesis tools induce on experimental evaluations and give answers to the question how convincing such evaluations can still be performed in such a setting.
\end{abstract}

\section{Introduction}

The problem of synthesizing reactive systems from linear-time temporal logic (LTL) specifications \cite{DBLP:conf/focs/Pnueli77,DBLP:conf/icalp/AbadiLW89,DBLP:conf/popl/PnueliR89,DBLP:journals/bsl/KupfermanV99,Vardi1995} has attracted many researchers in the past, leading to a tremendous amount of results in this area. Broadly, works contributing to the progress of its solution can be classified into two sorts. 

On the theory side, major breakthroughs have been obtained by establishing the 2EXPTIME-hardness of this problem \cite{DBLP:conf/icalp/PnueliR89,DBLP:conf/stoc/VardiS85,DBLP:journals/jcss/FischerL79}, along with asymptotically optimal automata-theoretic constructions for solving it \cite{DBLP:conf/popl/PnueliR89,DBLP:conf/icalp/AbadiLW89,Vardi1995}. Recent works are concerned with making these constructions easier \cite{DBLP:conf/focs/KupfermanV05,DBLP:conf/atva/ScheweF07a,DBLP:conf/vmcai/PitermanPS06} or enhancing the scope of the algorithms and hardness-results known to, e.g., distributed systems \cite{DBLP:conf/lics/KupfermanV01,DBLP:conf/lics/FinkbeinerS05}.

On the practical side, many works deal with the construction of sophisticated algorithms that aim at improving the scalability of current synthesis techniques \cite{DBLP:conf/fmcad/JobstmannB06,DBLP:conf/fmcad/SohailS09,DBLP:conf/cav/FiliotJR09,DBLP:conf/atva/FiliotJR10,DBLP:conf/vmcai/PitermanPS06,DBLP:conf/cav/Ehlers10}. While the 2EXPTIME-hardness of the LTL synthesis problem induces a limit on the effectiveness of any approach, this line of research is motivated by the observation that ``typical`` specifications in practice have a structure that can be exploited \cite{DBLP:conf/lics/Kupferman06,DBLP:conf/vmcai/SohailSR08,DBLP:journals/corr/abs-1006-1408,DBLP:conf/atva/FiliotJR10,DBLP:conf/fmcad/SohailS09,DBLP:conf/vmcai/PitermanPS06,DBLP:conf/cav/Ehlers10,DBLP:conf/cav/KupfermanPV06}.

While many works fall into both categories, contributions to the latter sort typically contain proofs of the usefulness of the proposed techniques obtained by experimentally evaluating a prototype implementation. This is commonly done by taking some example specifications (the so-called \newterm{benchmarks}) and showing that the prototype is able to handle these in reasonable time.

The situation is similar to the one for the problem of \newterm{satisfiability} (SAT) testing (see, e.g., \cite{Biere2009}), where despite its NP-completeness, an active area of research has emerged, witnessing its progress by the fact that
practical problems with millions of variables are nowadays routinely solved by modern SAT solvers. One of the key factors for this success is the possibility to perform meaningful benchmarking, which drives the development of new solution heuristics into those directions that appear to be most promising. Thousands of example problem instances can easily be used to obtain, optimise and test new approaches. As a result, the annual SAT solving competitions typically draw a lot of interest, for example 19 submissions to the main track in the 2010 SAT-Race \cite{SATRace2010}.

For the synthesis of reactive systems from linear-time specifications, however, there appears to be far less interest in tools. At the time of writing, to the best of our knowledge, there are only four publicly available synthesis tools\footnote{For the scope of this paper, we exclude all tools that aim at pure game solving, as here, the synthesis functionalities and the possibility to start from linear-time temporal logic is missing (which is typically not merely a preprocessing step). There is another tool called \RatsyCite, which we excluded as its synthesis functionality is mainly a reinterpretation of \Anzu{} (plus some preliminary implementation of the bounded synthesis approach \cite{DBLP:conf/atva/ScheweF07a,DBLP:conf/cav/FiliotJR09}) and the tool aims at providing an environment for specification engineering rather than being only a synthesis tool. Consequently, no experimental evaluation of the synthesis performance has been given in \cite{DBLP:conf/cav/BloemCGHKRSS10}. The \textsc{Jtlv} \cite{DBLP:conf/cav/PnueliSZ10} scripting environment that also has synthesis procedures has been excluded as no stand-alone synthesis tool exists and benchmarks comparisons are not available.},  namely \AnzuExtCite, \LilyCite, \AcaciaCite, and \UnbeastCite. Equally unfortunate, the amount of benchmarks available is rather low. One can identify (at least) three reasons for this state:

\begin{enumerate}
 \item A factor contributing to this difference is the fact that while rudimentary SAT solvers can be written in the order of hours, creating even a simple synthesis tool requires the implementation of comparably complex operations. As an example, for approaches working with deterministic automata, a construction similar to Safra's determinisation procedure \cite{phd-safra} needs to be performed, which has been argued to be notoriously complex to implement \cite{DBLP:conf/csl/HenzingerP06,DBLP:conf/tacas/HardingRS05,DBLP:conf/fmcad/JobstmannB06,DBLP:conf/lics/Kupferman06,DBLP:conf/cav/KupfermanPV06}. 

\item As a second reason, the publication schemes of the formal methods and SAT communities are different. While the latter appreciates work whose primary concern is to improve the scalability of current SAT solving techniques (see, e.g., \cite{DBLP:conf/sat/Kottler10,DBLP:conf/sat/NadelR10,POS2010} for recent such publications from 2010), works in the formal methods area are typically built around some appealing new concept, which is mostly only evaluated briefly on some prototype implementation (see, e.g., \cite{DBLP:conf/spin/GhafariHR10} for a pointer to a typical such a case in the area of software model checking, or simply compare recent practical synthesis papers \cite{DBLP:conf/atva/FiliotJR10,DBLP:conf/cav/Ehlers10,DBLP:conf/cav/FiliotJR09,DBLP:conf/fmcad/JobstmannB06,DBLP:conf/vmcai/PitermanPS06,DBLP:conf/fmcad/SohailS09,DBLP:conf/vmcai/SohailSR08}). 

Arguably, one of the main reasons for this difference is the fact that the roots of formal methods lie in theoretical computer science, where, given technical correctness and sufficient general style of writing, the main merits of a paper are seen in the significance of the conceptual contribution to the field \cite{DBLP:journals/iandc/Parberry94}. 
As a result, papers that propose improvements to current techniques that lack a major theoretical insight have a small chance of being accepted at major conferences even if the proposed techniques lead to significant speed-ups in the synthesis process. Consequently, time is typically only invested in writing synthesis tools after a new idea has been developed that is \emph{both} theoretically compelling and gives the impression that it will significantly improve upon the performance of previous techniques.

\item Third, even if time has been invested in writing a synthesis tool, a new technique still has to be shown to be competitive with earlier techniques. Typically, benchmarking is used for this purpose. In the scope of synthesis, however, it can be observed that this is by no means a trivial task -- all four currently available synthesis tools have different scopes and semantics. Also, the amount of benchmarks available is extremely low and the benchmarks in previous evaluations have often been rewritten to be compatible to the improvements proposed. This puts a high burden on the quality of future works in this area: comparability to previous works must be maintained in order to obtain a high credibility of the experimental evaluation. At the same time, as improvements to synthesis techniques often introduce additional details that must be taken care of in the evaluation (e.g., having two semi-algorithms running in parallel in \cite{DBLP:conf/cav/FiliotJR09,DBLP:conf/cav/Ehlers10} or the assumption dropping heuristic from \cite{DBLP:conf/atva/FiliotJR10}), it is very hard to produce an appealing evaluation without spending too much space in a publication on the details.
\end{enumerate}

Recently, the first of these problems has been attenuated by the availability of good LTL-to-Büchi translation tools \cite{DBLP:journals/sttt/RozierV10,DBLP:conf/cav/GastinO01,DBLP:conf/mascots/Duret-LutzP04} and determinisation and optimisation tools for $\omega$-automata \cite{DBLP:journals/tcs/KleinB06,DBLP:conf/sat/Ehlers10,DBLP:conf/spin/EhlersF10} on the one hand, and the availability of efficient binary decision diagram (BDD) libraries \cite{Somenzi98cudd:cu,DBLP:journals/tc/Bryant86} and satisfiability modulo theory (SMT) or SAT-solvers \cite{Biere2009} as reasoning backbones on the other hand. Thus, developers of new synthesis tools can build their implementations on top of such previous work. The second problem will hopefully lose impact over time when more interest in the practical side of reactive system synthesis is aroused.

In this paper, we approach the remaining third problem by giving both insights why experimental evaluations in the context of synthesis are notoriously harder than, e.g., in the SAT context, as well as proposing ''standardised`` evaluation schemes for synthesis tools that aim at simplifying further work in this area. We hope that our discussion helps interested observers of the advances in the practical approaches to synthesis (by providing a survey on the problem of evaluating synthesis tools, with special consideration of work already done in this area) as well as authors of future synthesis tools (by giving inspirations and in particular justification for the choice of their experimental settings) and
paper reviewers in the field (by explaining the difficulties of performing an experimental evaluation for synthesis tools).

We start by giving a definition of the LTL (open) synthesis problem in Section \ref{sec:Problem}. In Section \ref{sec:tools}, we review the synthesis approaches of the synthesis tools that were publicly available at the time of writing. Afterwards, we give some observations on these approaches (and their experimental evaluations). In Section \ref{sec:evaluationFramework}, we analyse the observations and propose a framework for future synthesis tool experimental evaluations. We conclude with a summary.

\section{The LTL open synthesis problem}
\label{sec:Problem}
We start by giving a problem description of reactive system synthesis that focusses on those aspects that require special attention when comparing synthesis approaches. 
Formally, a synthesis problem instance is a triple $\langle \AP_I, \AP_O, \psi \rangle$, where $\AP_I$ is a set of input atomic propositions, $\AP_O$ is a set of output atomic propositions, and $\psi$ is a formula in linear-time temporal logic (LTL) \cite{DBLP:conf/focs/Pnueli77} over $\AP_I \uplus \AP_O$. For the scope of this paper, we denote the LTL temporal operators for ''globally'', ''finally'' and ''next-time'' by $\mathsf{G}$, $\mathsf{F}$, and $\mathsf{X}$. For the ease of reading, we sometimes call the atomic propositions simply variables or bits.

We say that a triple $\langle \AP_I, \AP_O, \psi \rangle$ represents a \newterm{realisable} specification in the \newterm{Mealy-type semantics} if there exists some function $f : (2^{\AP_I})^+ \rightarrow 2^{\AP_O}$ such that for all $w = w_0 w_1 \ldots \in \AP_I^\omega$, we have $(w_0 \cup f(w_0)), (w_1 \cup f(w_0 w_1)), (w_2 \cup f(w_0 w_1 w_2)), \ldots \models \psi$. Likewise, we say that $\langle \AP_I, \AP_O, \psi \rangle$ is realisable in the \newterm{Moore-type semantics} if there exists some function $f : (2^{\AP_I})^* \rightarrow 2^{\AP_O}$ such that for all $w = w_0 w_1 \ldots \in \AP_I^\omega$, we have $(w_0 \cup f(\epsilon)), (w_1 \cup f(w_0)), (w_2 \cup f(w_0 w_1)), \ldots \models \psi$ for $\epsilon$ denoting the empty word. Specifications that are not realisable are called unrealisable in the respective semantics. 

Typically, realisability checking is performed by building a game between a \newterm{system player} and an \newterm{environment player}. In this setting, a function $f$ satisfying the constraints stated above is called a \newterm{winning strategy}. Details on the game-based view to synthesis can be found in \cite{DBLP:conf/dagstuhl/2001automata}.

It is well-known that whenever there exists some winning strategy for one of the semantics above, there also exists a finite representation of it. For the Mealy-type semantics, this representation is typically given as a Mealy automaton, whereas for the Moore-type semantics, Moore automata serve this purpose (see, e.g., \cite{Mueller2000}). 

Intuitively, the Mealy- and Moore-type semantics differ in the order of input and output. As an example, for the LTL formula $\psi = \mathsf{G}(r \leftrightarrow g)$, the specification $\langle \{r\}, \{g\}, \psi \rangle$ is realisable for the Mealy-type semantics but not for the Moore-type semantics. The reason is that in the Mealy-type semantics, the system already knows the input in the respective computation cycle when having to choose an output, whereas in the Moore-type semantics, the roles are swapped and thus the system has to guess whether $r$ is set or not when choosing whether $g$ should be set.

\section{Synthesis tools}
\label{sec:tools}
We briefly recapitulate the ideas behind the synthesis tools \AnzuCite, \LilyCite, \AcaciaCite, and \UnbeastCite. We use the terminologies from the respective papers and refer the reader not familiar with the terms used hereafter to them.

\subsection{\Anzu}
\label{ref:AnzuScopeDescription}
\paragraph{Scope:} This tool implements the concept of generalised reactivity(1) \cite{DBLP:conf/vmcai/PitermanPS06} synthesis (abbreviated as GR(1) synthesis) in the Mealy-type semantics. Here, the specification is restricted to be of the form $(a_1 \wedge a_2 \wedge \ldots \wedge a_n) \rightarrow (g_1 \wedge g_2 \wedge \ldots \wedge g_m)$ for some sets of assumptions $\{a_1, \ldots, a_n\}$ and guarantees $\{g_1, \ldots, g_m\}$. Every assumption is of one of the following forms:
\begin{enumerate}
 \item[(1)] $\psi_I$
 \item[(2)] $\mathsf{G}(\psi \rightarrow \mathsf{X}(\psi_I))$
 \item[(3)] $\mathsf{GF}(\psi)$
\end{enumerate}
where $\psi$ is an LTL-formula over $\AP_I \cup \AP_O$ free of temporal operators and $\psi_I$ is an LTL-formula over $\AP_I$ free of temporal operators . Likewise, all guarantees are of one of the following forms:
\begin{enumerate}
 \item[(1)] $\psi_O$
 \item[(2)] $\mathsf{G}(\psi \rightarrow \mathsf{X}(\psi_O))$
 \item[(3)] $\mathsf{GF}(\psi)$
\end{enumerate}
where $\psi_O$ is an LTL-formula over $\AP_O$ free of temporal operators. \Anzu{} requires the assumptions and guarantees to be in \newterm{Property Specification Language} (PSL) \cite{PSLBook} syntax and checks for \newterm{strong realisability} of the given overall specification, where unlike in normal realisability checking, safety guarantee violations are not tolerated in cases in which a safety assumption violation has not yet been witnessed but the system has a strategy to ensure that the overall input and output will not satisfy all assumptions (see, e.g., \cite{Klein2010}).
Specifications of the $(\bigwedge \text{assumptions}) \rightarrow (\bigwedge \text{guarantees})$ form, as used in \Anzu, typically occur in cases in which a part of a larger system is to be synthesized, where the set of assumptions represents the behaviour the part of the system to be synthesized can assume about the other parts of the system whereas the guarantees describe the requirements on the behaviour of the part of the system to be synthesized.

\paragraph{Techniques:} \Anzu{} implements generalised reactivity(1) synthesis \cite{DBLP:conf/vmcai/PitermanPS06} in a symbolic manner using binary decision diagrams (BDDs) \cite{DBLP:journals/tc/Bryant86,Somenzi98cudd:cu} as reasoning backbone. Here, a synthesis game is built whose state space consists of all variable valuations to the input and output atomic propositions. The LTL assumptions and guarantees of the forms $\psi_I$ and $\psi_O$ are encoded into the set of initial positions of the game, whereas assumptions and guarantees of the form $\mathsf{G}(\psi \rightarrow \mathsf{X}(\psi_I))$ and $\mathsf{G}(\psi \rightarrow \mathsf{X}(\psi_O))$ are encoded into its transition relation (describing the possible moves of the players). Then, a symbolic algorithm is used in which the system player tries to satisfy the specification  $(a'_1 \wedge a'_2 \wedge \ldots \wedge a'_{n'}) \rightarrow (g'_1 \wedge g'_2 \wedge \ldots \wedge g'_{m'})$ in this so-called \newterm{game arena} (consisting of the game positions and the transition relation), where $\{a'_1, \ldots, a'_{n'}\}$ are the assumptions of the form $\mathsf{GF}(\psi)$ and $\{g'_1, \ldots, g'_{m'}\}$ are the guarantees of the form $\mathsf{GF}(\psi)$. Implementations are extracted by building circuits for computing the outputs from the BDD representation of the winning state set while performing a care-set optimisation to the BDD after every step \cite{DBLP:conf/date/BloemGJPPW07,DBLP:journals/entcs/BloemGJPPW07}.

\paragraph{Experimental evaluation:} The usefulness of the tool \Anzu{} has been shown on two  case studies: an AMBA AHB arbiter \cite{AMBA1999} specification (simplified by leaving out bus splits and early burst terminations) and a generalised Buffer (GenBuf) controller that has been described by IBM for tutorial purposes \cite{IBMTutorial}. Both case studies contain several assumptions and guarantees and are scalable by the numbers of clients.

\subsection{\Lily}

\paragraph{Scope:} The tool \Lily{} accepts arbitrary LTL formulas in PSL syntax as specifications. If multiple LTL formulas are found in the input file, they are treated in a conjunctive manner, except for those that are preceded by an \texttt{assume} keyword, which are used as assumptions for the overall specification.

\paragraph{Techniques:} \Lily{} implements the optimisations to the Safraless synthesis \cite{DBLP:conf/focs/KupfermanV05} approach presented in \cite{DBLP:conf/fmcad/JobstmannB06}. In the first step, the negation of the specification is converted to a node-labelled nondeterministic Büchi word (NBW) automaton using the LTL-to-Büchi translator \WringCite. The Büchi automaton is then converted  to a universal co-Büchi tree automaton (UCT) that checks for the satisfaction of the original specification along all computation tree paths, which is in turn tested for emptiness using a construction proposed by Kupferman and Vardi \cite{DBLP:conf/focs/KupfermanV05}, utilising alternating weak tree automata (AWT) and non-deterministic Büchi tree (NBT) automata. Jobstmann and Bloem \cite{DBLP:conf/fmcad/JobstmannB06} add various optimisations to these steps. 
The conversion of the UCT to the AWT is parametrised by some constant $k$, which influences the size of the NBT produced and thus the running time of the overall algorithm. While low values for $k$ typically suffice for realisable specifications in practice, a relatively large value of $k$, exponential in the size of the NBW (or, alternatively, doubly-exponential in the length of the LTL specification), is needed to have the emptiness of the language of the NBT imply the emptiness of the UCT language (except if the UCT turns out to be weak), and thus prove unrealisability of the given specification. 
To avoid this problem, \Lily{} can also be run in a special unrealisability detection mode. In this case, it checks the realisability of the negated specification with swapped inputs and outputs (and a slight modification of the resulting specification to convert its Mealy-type semantics to Moore-type again). Then, for unrealisable specifications, only a small value of $k$ is typically needed in practice to identify them as such.

\paragraph{Experimental evaluation:} In \cite{DBLP:conf/fmcad/JobstmannB06}, the performance of \Lily{} is evaluated on some specifications written by the authors of that paper, representing mostly arbiter variations and traffic light controllers. The evaluation is focussed on proving that the optimisations proposed in that paper contribute significantly to having low running times of the tool.

\subsection{\Acacia}

\paragraph{Scope:} The tool \Acacia{} has the same input syntax as \Lily{} and also uses Moore-type semantics. There exist two versions of \Acacia{}. \Acacia{} 2009 implements the techniques described in \cite{DBLP:conf/cav/FiliotJR09}, while \Acacia{} 2010 also implements those of \cite{DBLP:conf/atva/FiliotJR10}. The latter version includes support for making assumptions local to some set of guarantees. The specification then consists of a conjunction of sub-specifications of the form $(\bigwedge \text{assumptions}) \rightarrow (\bigwedge \text{guarantees})$. All assumptions and guarantees of the conjuncts are assumed to be given separately.

\paragraph{Techniques:} \Acacia{} is based on the concept of bounded synthesis \cite{DBLP:conf/atva/ScheweF07a,DBLP:conf/cav/FiliotJR09}, a refinement of the Safraless synthesis techniques proposed in  \cite{DBLP:conf/focs/KupfermanV05}. Here, as in \Lily{}, the specification is first negated and then converted to a node-labelled Büchi automaton. As \Lily, \Acacia{} uses \WringCite{} for this purpose. Afterwards, the Büchi automaton is converted to a universal co-Büchi tree automaton (UCT)  that checks for the satisfaction of the specification along all paths of a computation tree. This UCT is used as a basis for building a series of synthesis safety games, where for a successively increasing \newterm{bound value} $k$, for every state $q$ in the UCT, the maximum number of visits to rejecting states in the UCT from its initial state along some path to $q$ for the input/output played in the game so far is encoded into the game positions. Once one of these counters exceeds the value $k$, the game is lost for the system player. 
The main idea of \Acacia{} is to use anti-chains as an efficient representation of the \newterm{frontier sets} (i.e., pre-fixed points of winning positions) occurring during the safety game solving process. This representation makes use of the fact that the set of possible future behaviours of the system player in a position $p_1$ can only be larger than when being in a game position $p_2$ if all counters in $p_1$ are less than or equal to those in $p_2$. Thus by storing only states whose counter vectors are not \newterm{dominated} by the counter vectors of other states in the pre-fixed point during the solving process, redundancies can be avoided. 

As in \Lily, the value of $k$ required to conclude the unrealisability of a specification is exponential in the size of the UCT or doubly-exponential in the length of the LTL specification. The authors thus propose to run \Acacia{} two times in parallel, where in the first run, realisability is checked and in the second run, unrealisability is tested. As in \Lily, in the latter case, the specification is negated, a conversion between Mealy-type and Moore-type semantics takes place, and the inputs and outputs are swapped.

\Acacia 2010 adds some additional features. Here, the game solving process is made compositional. Recall that in \Acacia{} 2010, the specification is supposed to consist of a conjunction of sub-specifications of the form $(\bigwedge \text{assumptions}) \rightarrow (\bigwedge \text{guarantees})$. In this setting, the safety games for the synthesis process can be built separately, preliminarily solved independently and finally composed on-the-fly during the solving process for the game representing the overall specification. \Acacia{} 2010 furthermore adds the possibility to use the \textsc{OTFUR} mixed forward-backward game solving algorithm \cite{DBLP:conf/icalp/LiuS98,DBLP:conf/concur/CassezDFLL05} instead of the classical backward safety game solving algorithm. Finally, for cases in which the specification is only a single formula of the form $(\bigwedge \text{assumptions}) \rightarrow (\bigwedge \text{guarantees})$, \Acacia{} can rewrite this specification into the form $\bigwedge_{g \in \text{guarantees}}\left((\bigwedge \text{assumptions}) \rightarrow g \right)$ in order to benefit from the compositional algorithms implemented. In this case, an assumption dropping heuristic is used to remove some assumption copies in this formula, which reduces the problem that the assumptions are replicated for all guarantees in this setting. Using the heuristic makes the approach however incomplete.

\paragraph{Experimental evaluation:} In \cite{DBLP:conf/cav/FiliotJR09}, the focus of the experimental evaluation lies on proving that \Acacia{} improves upon the performance of \Lily{}, using the fact that the semantics are compatible. The authors of \cite{DBLP:conf/cav/FiliotJR09} show that on the examples from \cite{DBLP:conf/fmcad/JobstmannB06}, using the anti-chains approach typically results in lower computation times and that the Büchi automaton building time surprisingly dominates the overall synthesis time. One of the example specifications is made scalable and it is proven that the anti-chains approach is much faster here. Another set of variations of one of the \Lily{} examples is used as a further benchmark set.

In \cite{DBLP:conf/atva/FiliotJR10}, the 2010 version of \Acacia{} is evaluated with several different choices for (1) whether game solving should be performed backwards or in a forward-backward manner, (2) whether monolithic or compositional synthesis should be performed, and (3) whether the assumption dropping heuristic should be used (only in the compositional case). Apart from the benchmarks also used in \cite{DBLP:conf/cav/FiliotJR09}, the generalised Buffer (GenBuf) controller \cite{IBMTutorial} specification that was also used for benchmarking \Anzu{} has been formulated in a way such that the assumptions to the environment are local to some sets of guarantees, such that compositional synthesis can be performed directly. This benchmark is used to show the benefits of the compositional approach.

\subsection{\Unbeast}

\paragraph{Scope:} The tool \UnbeastCite{} focusses on specifications of the form $(\bigwedge \text{assumptions}) \rightarrow (\bigwedge \text{guarantees})$ and uses Mealy-type semantics. By using an input language based on XML, incorrect presumptions by the user about precedences of temporal operators in LTL are avoided. The assumptions and guarantees are given separately in the XML input file.

\paragraph{Techniques:} The \Unbeast{} tool implements the synthesis techniques presented in \cite{DBLP:conf/atva/ScheweF07a,DBLP:conf/cav/Ehlers10}. The library \CUDD{} \cite{Somenzi98cudd:cu} is used for constructing and manipulating BDDs during the synthesis process. 

The first step is to determine which of the given assumptions and guarantees are safety formulas. In order to detect also simple cases of \newterm{pathological safety} \cite{DBLP:journals/fmsd/KupfermanV01}, this is done by computing an equivalent Büchi automaton using an external LTL-to-Büchi converter such as \LTLtoBACite{} or \SpotCite, and examining whether all maximal strongly connected components in the computed automaton do not have infinite non-accepting paths. Special care is taken of so-called of \newterm{bounded look-ahead safety formulas}.

In a second step, for the set of bounded look-ahead assumptions and the set of such guarantees, safety automata for their respective conjunctions are built. Both of them are represented in a symbolic way using BDDs. For the remaining safety assumptions and guarantees, safety automata are built by taking the Büchi automata computed in the previous step and applying a subset construction for determinisation in a symbolic manner. For the remaining non-safety parts of the specification, a combined universal co-Büchi automaton is computed by calling the external LTL-to-Büchi tool again. 

In the next phase, the given specification is checked for realisability. This is done almost as in \Acacia{} 2009, i.e., for a successively increasing so-called \textit{bound value}, the bounded synthesis approach \cite{DBLP:conf/atva/ScheweF07a,DBLP:conf/cav/FiliotJR09} is performed by building a safety automaton from the co-Büchi automaton for the non-safety part of the specification and solving the safety games induced by a special product of the automata involved \cite{DBLP:conf/cav/Ehlers10}. However, instead of anti-chains, BDDs are used. 

Finally, if the specification is found to be realisable (i.e., the game computed in the previous phase is winning for the player representing the system to be synthesised), the symbolic representation of the winning states of the system is used to compute a prototype implementation satisfying the specification in a fully symbolic way, using a slight simplification of the algorithm from \cite{DBLP:conf/cav/KukulaS00}. However, the implementations generated are typically relatively large. As \Acacia{}, to also detect unrealisable specifications, \Unbeast{} needs to be run two times in parallel.

\paragraph{Experimental evaluation:} \Unbeast{} was evaluated on the specifications defined in \cite{DBLP:conf/fmcad/JobstmannB06} as well as on those given in  \cite{DBLP:conf/cav/FiliotJR09}. The Moore-type semantics from these examples have been adapted to the Mealy-type semantics of \Unbeast{} by prefixing all occurrences of input atomic propositions with an LTL next-time operator (see, e.g., \cite{DBLP:conf/fmcad/JobstmannB06}).

Additionally, a scalable load balancing case study is presented in \cite{DBLP:conf/cav/Ehlers10}, having a Mealy-type semantics. For comparison and usage with \Acacia{} and \Lily{}, the examples have been transformed to Moore-type semantics by prefixing all occurrences of output atomic propositions with an LTL next-time operator.

\section{Observations on the differences and similarities of the synthesis tools}
\label{sec:observations}

We continue with a discussion of the similarities and differences between the synthesis tools. The arguments to follow form the foundation of the experimental evaluation frameworks proposed in Section~\ref{sec:evaluationFramework}.

\subsection{The incomparable scopes \& semantics of the tools}

When comparing the tools considered in this paper, it is striking that all four tools have incompatible specification languages. Only \Acacia{} 2009 and \Lily{} have the same input specification format (however, \Acacia{} 2010 adds local assumptions to the input language which cannot be interpreted correctly by \Lily). 

It is fair to raise the question why this is the case, given the fact that the number of benchmark sets for synthesis is rather low, so one would expect that the scopes and semantics are compatible in order to have as many benchmarks available as possible.
In this paper, we conjecture that the reason for this situation is that \emph{the scopes of the tools are strongly adapted to the techniques implemented}, but whenever the choice does not matter, as close to the literature as possible.

We begin our discussion of this observation with the tool \Anzu. The generalised reactivity(1) synthesis technique is implementable in both Mealy and Moore semantics. The tool \Anzu{} uses Mealy semantics, as the description of the synthesis algorithm in \cite{DBLP:conf/vmcai/PitermanPS06}. The assumptions and guarantees allowed are precisely those that can be processed by the algorithm without giving up the idea that the state space of the underlying game is the set of input and output variable valuations. This can be seen from the fact that from the formula types given in Section \ref{ref:AnzuScopeDescription}, type (1) is only an initial state condition, and type (2) can be encoded into the transition relation of a game with such a structure. Formulas of type (3) are precisely those that can then be given to the actual solving process as liveness parameters.

In contrast to \Anzu, \Lily{} uses Moore-type semantics and some PSL-like input file syntax. Since \Lily{} bases on tree automaton techniques, this is not surprising: using a Mealy-type semantics in the context of tree automata would require that the labelling of the initial node of a computation tree (that is either accepted or rejected by the tree automaton) is ignored, as the node labels represent the output of the system. On a theoretical level, such a definition would look unnecessarily awkward, which is why the Moore-type semantics are usually preferred in this context.

As \Lily, \Acacia{} uses a Moore-type semantics and the same syntax as \Lily{}. According to \cite{DBLP:conf/cav/FiliotJR09}, the authors wanted to keep \Acacia{} 2009 comparable to \Lily{}. An additional reason for keeping the Moore-type semantics is the better applicability of the results: specifications that are found to be realisable in the Moore-type semantics are also realisable in the Mealy-type semantics, but not vice versa. Only in \Acacia{} 2010, the input language is extended in order to accommodate the new features proposed in \cite{DBLP:conf/atva/FiliotJR10}.

\Unbeast, on the other hand, uses a Mealy-type semantics and has its own XML-based input file format. In \cite{DBLP:conf/cav/Ehlers10}, it has been argued that specifications often become shorter and thus the (edge-labelled) Büchi automata become smaller in the Mealy setting, which is beneficial for a BDD-based approach. For example, when specifying some immediate output consequences of some input such as $\mathsf{G}(r \rightarrow g)$ for some input set $\{r\}$ and output set $\{g\}$, taking the Moore semantics would require the introduction of a next-time operator into the formula, which would be reflected in the automaton size. The XML-based input language has been used in order to circumvent the necessity for a complicated formula parser, but also to make the operator precedences explicit. 

\subsection{Comparability of the examples}
\label{sec:ComparabilityOfExamples}

The arguments from the preceding subsection explain why the scopes and semantics of the tools are different. Nevertheless, it does not explain why different specification sets have been used, as benchmarks for \Anzu{} could be converted to benchmarks for the other tools and the conversion between Mealy- and Moore-type semantics is rather simple. Still, the only case in which benchmarks were converted for an experimental evaluation was in \cite{DBLP:conf/cav/Ehlers10}, where \Lily's and \Acacia's examples were used for evaluating \Unbeast{} (and vice versa). In some cases, benchmarks have been rewritten, e.g., the IBM generalised buffer specification, which was used to evaluate \Anzu{}, has been altered in \cite{DBLP:conf/atva/FiliotJR10} to a form in which the assumptions were made local. 
Also, there are two other publications \cite{DBLP:conf/fmcad/SohailS09,Morg10} reporting on experimental results for synthesis approaches using generalized parity games. In both of them, the feasibility of their approaches is shown using different reformulations of the AMBA arbiter example that was also used for \Anzu{}. 

An explanation for this fact was given in \cite{DBLP:conf/vmcai/SohailSR08,DBLP:conf/fmcad/SohailS09,DBLP:conf/cav/BloemCGHJ10}. In fact, the GR(1) synthesis approach implemented in \Anzu{} can accommodate all types of assumptions and guarantees that are representable as deterministic Büchi automata (DBA) \cite{DBLP:conf/cav/BloemCGHJ10,DBLP:conf/sat/Ehlers10}. In order to fit into the input language of \Anzu, however, the output bit set of the system to be synthesized has to be extended by state bits of the automaton. Somenzi and Sohail coined the term ``pre-synthesis'' for such an encoding, as converting an assumption or guarantee to such an automaton and encoding it into some output bits in a good way is a problem on its own for BDD-based techniques (see, e.g., \cite{DBLP:journals/amcs/GostiVSS07,DBLP:conf/aspdac/ForthM00}). Thus, a lot of effort has been put into a good reformulation of the problem description before checking realisability. An equivalent approach to using DBAs is to introduce so-called \newterm{auxiliary signals} (or auxiliary variables) into the design \cite{DBLP:journals/corr/abs-1001-2811,DBLP:journals/entcs/BloemGJPPW07,DBLP:conf/date/BloemGJPPW07}. It has been noted that rewriting a specification using different signals can significantly speed up the synthesis process \cite{DBLP:journals/corr/abs-1001-2811,DBLP:journals/entcs/BloemGJPPW07} and for the AMBA AHB specification, this has also been done. As a consequence, it is not surprising that a specification for which pre-synthesis was performed and that has been optimized towards \Anzu, neither the authors of \Acacia{} nor \Unbeast{} (nor the authors of the works using generalized parity automata \cite{DBLP:conf/fmcad/SohailS09,Morg10}) used the AMBA AHB arbiter specifications in the form provided with the \Anzu{} tool for comparisons.

With respect to the fact that the IBM Generalised Buffer example has been altered for usage with \Acacia{} 2010, the situation is similar: in the original specification, the assumptions were not localised; defining the scope of the assumptions was simply not an issue in this case. As soon as techniques are introduced that can make use of such local assumptions, the situation changes. 

\subsection{Complexity of the workflows}

Except for \Anzu{}, the workflows, i.e., the numbers and orders of computation steps in the realisability checking process, of the tools discussed here are rather complicated. \Lily{} and \Acacia{} 2009 first convert the specification to a universal Büchi automaton and then perform, for some successively increasing bound value, a realisability check over this automaton. In order to also detect unrealisable specifications, the check must additionally be ran for the negated specification with a conversion between the two semantics types in parallel. The workflow of \Unbeast{} is similar. In contrast to many other formal methods experimental evaluations, this whole process is relatively complicated and might easily appear less compelling than more simple schemes that are used in, for example, SAT solvers.

It is fair to conjecture that future workflows will even be more complicated. Take for example, \Anzu{}, which has a relatively straight-forward workflow. As it has been argued that the generalised reactivity(1) synthesis approach that is used in this tool could handle all assumptions and guarantees that are representable by deterministic Büchi automata, developing a preprocessor that takes LTL specifications of this kind and produces equivalent \Anzu{} specifications appears to be worthwhile to write. However, such a preprocessor would have a very complicated workflow. After converting the assumptions and guarantees to Büchi automata, these have to be determinised (whenever possible), using an external tool like \textsc{ltl2dstar} \cite{DBLP:journals/tcs/KleinB06}. Afterwards, it is possibly wise to try some exhaustive minimisation method for these automata \cite{DBLP:conf/sat/Ehlers10}. Then, the automata also have to be symbolically encoded \cite{DBLP:journals/amcs/GostiVSS07,DBLP:conf/aspdac/ForthM00}. Furthermore, the time spent on optimising the automata has to be balanced against the overall computation time in order to avoid running out of time in the automaton optimisation step.\footnote{In benchmark comparisons, it is customary to restrict the running times of the tools. Such a time restriction is the typical answer to the problem that in most experimental evaluations, there are some benchmark/tool combinations that do not yield a result even after days or weeks of computation time.} All in all, these aspects make the whole synthesis process quite complicated and arguably, less compelling than other approaches, which ultimately reduces the publishability of any result on such a workflow, which in turn leads to little incentive to perform research or write tools in this area.

\section{Providing a framework for future evaluations}
\label{sec:evaluationFramework}

The preceding sections discussed the difficulties of composing meaningful experimental evaluations of synthesis tools. Nevertheless, as benchmarking is often considered to be the only way to distinguish promising ideas from the ones that are likely not to be useful (see, e.g., \cite{DBLP:journals/computer/Tichy98}), in this section, we propose \emph{three evaluation schemes} for each of the problems of \emph{using appropriate benchmarks} and \emph{dealing with complex workflows} whose compositions respect the difficulties discussed earlier. The schemes are ordered from the minimum requirement to show that a new technique is worthwhile considering to the ``superior'' scheme that demonstrates clear advantages over previous techniques.

\subsection{Benchmarking}
\subsubsection{Comparison using the home field advantage}

It is fair to say that a new approach should beat older approaches at least in the cases in which it has a natural advantage. This is typically shown by taking some example specification that falls into the class of systems the new approach is intended to be applied to, applying a prototype implementation of the approach to it, and showing that previous tools perform worse using an automatic, ingenuine, conversion to the semantics/scopes of the previous tools. This means in particular to convert between Mealy- and Moore-type semantics if applicable. Competitor tools which can only handle a subset of the language of the new prototype tool need not be considered. 

\subsubsection{Comparison from a neutral view-point}

One problem of benchmarking tools with different scopes and semantics against the same examples is that \emph{specifications are typically geared towards the usage with a certain tool}. A typical example is the pre-synthesis process discussed in Section \ref{sec:ComparabilityOfExamples} that ensures that the specification of a system falls into the class handled by the GR(1) synthesis tools. After this has been done, the specification is not only suitable but also optimised for such a tool. As many signalling bits are introduced in the process, tools like \Lily{} and \Acacia{} 2009 that are explicit in the input and output bit valuations have problems with handling such pre-synthesized specifications even in cases in which they can deal with the non-pre-synthesized versions.

A similar situation arises for example when localising the assumptions (as discussed in Section \ref{sec:ComparabilityOfExamples}): doing so is beneficial for \Acacia{} 2010, but renders the optimisations of \Unbeast{} unusable as the input is then no longer in the $(\bigwedge \text{assumptions}) \rightarrow (\bigwedge \text{guarantees})$ form. 

As a solution to this problem, we propose the following scheme: given a setting, the specification is written for all tools to be compared individually, taking care of their specialities. If the prototype tool of a new approach performs better in such a situation  than previous tools, it is clear that the techniques proposed have their merits if used correctly when modelling a specification. It should be noted, however, that this scheme favours tools that require some form of pre-synthesis: by rewriting the specification for the simpler tool in a smart way, its performance can often greatly be increased. As an example, we refer to the work on rewriting the AMBA AHB bus arbiter specification  \cite{DBLP:journals/corr/abs-1001-2811}.

\subsubsection{Beating the other tools where they have a natural advantage}
As a third scheme, we propose that if a prototype implementation of some approach can beat other tools on benchmark suites on which they have a natural advantage, this should suffice to show the merits of a new approach without doubt. In order to do so, one would typically use an automatic converter between the scope and semantics (if necessary) of the other tool and the scope and semantics of the new prototype tool to import benchmarks originally written for the other tool. The converter must not apply sophisticated optimisations on the specification. As an example, converting the LTL formula $\mathsf{G}\mathsf{F} \mathsf{X} p$ to $\mathsf{G}\mathsf{F} p$ for some atomic proposition $p$ during the adaptation of the Mealy/Moore-type semantics should be considered to be fair, whereas rewriting a guarantee into a simpler one that is only equivalent if the given assumptions also hold is probably too complex for this scheme.

\subsection{Complex workflows}
\subsubsection{Basic scheme}
In order to combat the problem of having workflows that involve multiple steps that can be be skipped without obstructing the steps following (like for example automaton optimisations), we propose the following scheme: for successively increasing timeout values (using a reasonable granularity), the synthesis approach is performed using the given timeout value for \emph{all} individual sub-steps involved until the respective tool execution yields an answer. If the least timeout value that leads to a result for the new technique is lower than the least such timeout value for previous approaches, it is shown that the new approach has some merits. 

\subsubsection{Advanced scheme}

As an extension to the basic scheme, it is worthwhile to show that the timeout value obtained in the basic scheme does not make the old approaches look bad unnecessarily. Let $A$ be the least timeout value (for every step of the workflow) tried such that the prototype tool of the new approach terminates with an answer. Let $B$ be the overall running time of the process. If it can be shown that the other approaches do not even terminate with a timeout of $B$ for each step, additional justification for the new approach is obtained. Of course, many intermediate variations between the basic and advanced schemes are possible.

\subsubsection{Simple scheme}

Probably the most convincing way to solve the problem of having complex workflows is to set static timeouts for the individual steps of the workflow and to just measure the overall running time. If it is better than those of other tools, this clearly shows the efficiency of the new approach. Obviously, this scheme is hard to follow when comparing against an other approach that has itself many steps which introduce a need for individual timeouts if the author of the other tool has not provided good values for these. Additionally, special care must be taken not to ``overfit`` \cite{DBLP:journals/heuristics/Falkenauer98} the timeouts for the individual steps -- tuning these values for the new prototype tool against a benchmark set and then evaluating on the same set against the other tools in a publication is not fair and can be considered to be scientifically unsound. 

\section{Conclusion}

In this paper, we discussed the problems of experimentally evaluating a synthesis tool. We discussed three major issues: the incomparable semantics and scopes of the tools, the bad comparability of the tools with respect to the benchmarks available and the complexities of the workflows. Three evaluation schemes to combat the first two of these problems and three schemes to account for the complex workflows have been presented. 

While the workflow evaluation schemes cannot fully remove the problem that experimental evaluations using these are often not fully compelling to the reader of a scientific publication, they introduce means of comparing tools if they have parts in their workflow that may time out without prohibiting later steps (like, e.g., automaton optimisation). We must admit that currently, there is no synthesis tool that performs such steps. However, this on its own is an interesting fact: due to the immense set of techniques proposed in the literature for reducing the sizes and numbers of automata representing the overall specification, operations such as determining whether a guarantee is actually necessary in a specification or more complicated automaton minimisation techniques are not used in current tools yet even though the theory behind these operations has been established \cite{DBLP:conf/icalp/GreimelBJV08,DBLP:conf/sat/Ehlers10,DBLP:conf/spin/EhlersF10,DBLP:conf/concur/ClementeM10}. Thus, we hope that the three workflow evaluation schemes proposed help to level the way to further advances in this area.

As a final note, we would like to defend the argumentation in this paper against the point of view that establishing a common file format with its clearly defined semantics and scope and requiring all future tools to use it as a basis is a way to fight the benchmarking problem. We have shown in Section \ref{sec:observations} that the choice of techniques affects the choice of the semantics of a tool. Thus, picking one particular scope and semantics would drive the evolution of the tools and thus also the theory into a certain direction while ignoring possibilities apart from techniques not suitable for the scope and semantics agreed upon. However, as even if leaving this consideration apart, the form of a ''typical'' specification in practice (conjunction of guarantees \cite{DBLP:conf/atva/FiliotJR10,DBLP:conf/cav/KupfermanPV06,Morg10} vs.\ $(\bigwedge \text{assumptions}) \rightarrow (\bigwedge \text{guarantees})$ form \cite{DBLP:conf/date/BloemGJPPW07,DBLP:conf/cav/Ehlers10,DBLP:journals/entcs/BloemGJPPW07,DBLP:conf/cav/BloemCGHJ10,DBLP:journals/corr/abs-1001-2811}) is not agreed upon, it is fair to argue that consensus will not be reached within the next few years. Also, an implicit or explicit requirement that tools with a complex workflow should always be evaluated in a way similar to the simple scheme proposed here is highly problematic: due to the low number of meaningful benchmarks, fixing good values as timeouts for the intermediate steps without overfitting for the concrete set of benchmarks in the evaluation is hardly possible. As a result, such a requirement would basically rule out complex optimisations a-priori (or require cheating by an author in the paper by overfitting), which is highly questionable for practical approaches to a problem that is, after all, still 2EXPTIME-complete.

\section*{Acknowledgements}

This work was supported by the German Research Foundation (DFG)
as part of the Transregional Collaborative Research Center
``Automatic Verification and Analysis of Complex Systems''
(SFB/TR 14 AVACS).
The author wants to thank Barbara Jobstmann and Emmanuel Filiot for helpful comments on the descriptions of the techniques employed in \Anzu{}, \Lily{} and \Acacia.

\bibliographystyle{eptcs}
\bibliography{bib}

\end{document}